\documentstyle[multicol,aps,prl]{revtex} 

\renewcommand{\narrowtext}{\begin{multicols}{2} \global\columnwidth20.5pc}
\renewcommand{\widetext}{\end{multicols} \global\columnwidth42.5pc}
\def\al{\alpha}
\def\be{\beta}
\def\ga{\gamma}
\def\de{\delta}
\def\ep{\epsilon}
\def\ve{\varepsilon}

\def\ka{\kappa}
\def\la{\lambda}

\def\si{\sigma}

\def\ph{\phi}

\def\ch{\chi}
\def\ps{\psi}
\def\om{\omega}

\def\De{\Delta}

\def\Si{\Sigma}

\def\Om{\Omega}

\def\cl{{\cal L}}
\def\cE{{\cal E}}
\def\fr#1#2{{{#1} \over {#2}}}

\def\half{{\textstyle{1\over 2}}}
\def\frac#1#2{{\textstyle{{#1}\over {#2}}}}

\def\lsim{\mathrel{\rlap{\lower4pt\hbox{\hskip1pt$\sim$}}
    \raise1pt\hbox{$<$}}}
\def\gsim{\mathrel{\rlap{\lower4pt\hbox{\hskip1pt$\sim$}}
    \raise1pt\hbox{$>$}}}
\def\sqr#1#2{{\vcenter{\vbox{\hrule height.#2pt
         \hbox{\vrule width.#2pt height#1pt \kern#1pt
         \vrule width.#2pt}
         \hrule height.#2pt}}}}

\def\lrpartial{\raise 1pt\hbox{$\stackrel\leftrightarrow\partial$}}

\def\lrdla{\stackrel{\leftrightarrow}{D^\la}}

\newcommand{\beq}{\begin{equation}}
\newcommand{\eeq}{\end{equation}}
\newcommand{\bea}{\begin{eqnarray}}
\newcommand{\eea}{\end{eqnarray}}
\newcommand{\rf}[1]{(\ref{#1})}
 
\begin{document}

\title{CPT and Lorentz tests with muons}

\author{Robert Bluhm,$^a$ V.\ Alan Kosteleck\'y,$^b$  
and Charles D.\ Lane$^b$} 

\address{$^a$Physics Department, Colby College, Waterville, ME 04901} 
\address{$^b$Physics Department, Indiana University, Bloomington, 
IN 47405} 

\date{IUHET 410, COLBY-99-03; 
accepted for publication in Phys.\ Rev.\ Lett.} 

\maketitle

\begin{abstract}

Precision experiments with muons are sensitive
to Planck-scale CPT and Lorentz violation
that is undetectable in other tests.
Existing data on the muonium ground-state hyperfine structure 
and on the muon anomalous magnetic moment 
could be analyzed to provide dimensionless figures of merit 
for CPT and Lorentz violation
at the levels of $4\times 10^{-21}$ and $10^{-23}$.

\end{abstract}

\narrowtext

The minimal standard model of particle physics 
is CPT and Lorentz invariant. 
However,
spontaneous breaking of these symmetries may occur
in a more fundamental theory incorporating gravity 
\cite{kskp,cpt98}.
Minuscule low-energy signals of CPT and Lorentz breaking 
could then emerge in experiments sensitive to effects 
suppressed by the ratio 
of a low-energy scale to the Planck scale.
At presently attainable energies,
the resulting effects would be described 
by a general standard-model extension 
\cite{ck}
that allows for CPT and Lorentz violation
but otherwise maintains conventional properties
of quantum field theory,
including gauge invariance, renormalizability,
and energy conservation.

In the present work,
we study the sensitivity 
of different muon experiments
to CPT and Lorentz violation.
Planck-scale sensitivity to possible effects
is known to be attainable 
in certain experiments without muons.
These include,
for example,
tests with neutral-meson oscillations
\cite{mesontests,ckpv},
searches for cosmic birefringence
\cite{cfj,ck,jk99},
clock-comparison experiments
\cite{cctests,kl99},
comparisons of particles and antiparticles in Penning traps
\cite{penningtests,bkr9798},
spectroscopic comparisons of hydrogen and antihydrogen
\cite{bkr99},
measurements of the baryon asymmetry 
\cite{bckp},
and observations of high-energy cosmic rays
\cite{cg99}.
However,
in the context of the standard-model extension,
dominant effects in the muon sector
would be disjoint from those in any of the above experiments 
because the latter involve only photons, hadrons, and electrons.
Moreover,
if the size of CPT and Lorentz violation scales with mass,
high-precision experiments with muons would represent 
a particularly promising approach to detecting 
lepton-sector effects from the Planck scale.

The standard CPT test involving muons 
compares the $g$ factors for $\mu^-$ and $\mu^+$,
with a bound 
\cite{cernmug,farley}
given by the figure of merit
\beq
r_g^\mu \equiv |g_{\mu^+}-g_{\mu^-}|/g_{\rm av} \lsim 10^{-8}
\quad .
\label{rg}
\eeq
We show here that data from experiments normally not associated
with CPT or Lorentz tests,
including
muonium microwave spectroscopy 
\cite{muonium99}
and $g-2$ experiments 
on $\mu^+$ alone
\cite{muong99},
can indeed provide Planck-scale sensitivity
to CPT and Lorentz violation.

For the experiments considered here,
it suffices to consider 
a quantum-electrodynamics limit of the standard-model extension
incorporating only muons, electrons, and photons.
Other terms in the full standard-model extension
would be irrelevant or lead only to subdominant effects.
In natural units with $\hbar = c = 1$,
the Lorentz-violating lagrangian terms of interest are
\bea
\cl &=&
- a_{\ka \, AB} \, \bar l_A \ga^\ka l_B
- b_{\ka \, AB} \, \bar l_A \ga_5 \ga^\ka l_B
  \nonumber \\
&&
- \half H_{\ka\la \, AB} \,
  \bar l_A \si^{\ka\la} \l_B
+ \half i c_{\ka\la \, AB} \,
  \bar l_A \ga^\ka \lrdla \l_B
  \nonumber \\
&&
+ \half i d_{\ka\la \, AB} \,
  \bar l_A \ga_5 \ga^\ka \lrdla \l_B
\quad .
\label{lqed}
\eea
Here,
the lepton fields are denoted by $l_A$
with $A=1,2$ corresponding to
$e^-$, $\mu^-$,
respectively,
and $i D_\la \equiv i \partial_\la - q A_\la$
with charge $q = - |e|$.
To avoid confusion with 4-vector indices,
the symbol $\mu$ is reserved in this work 
solely as a label for the muon.

The terms associated with the parameters
$a_{\ka \, AB}$, $b_{\ka \, AB}$
are CPT odd,
while the others are CPT even. 
All the parameters in Eq.\ \rf{lqed} are assumed small,
and they all are hermitian 2$\times$2 matrices in flavor space.
For example,
\beq
b_{\ka} = 
\left( \matrix{b_\ka^e &b_\ka^{e\mu} \cr 
b_\ka^{\mu e} &b_\ka^\mu \cr} \right)
\quad ,
\label{amatrix}
\eeq
where $b_\ka^e$, $b_\ka^\mu$ are associated with terms 
preserving lepton number while 
the others are associated with terms violating it.
Since the usual standard model conserves lepton number,
leading-order rates for processes 
that violate lepton number 
in the standard-model extension 
must be quadratic in the flavor-nondiagonal parameters
$b_\ka^{e \mu}$, etc. 
In contrast,
processes violating Lorentz symmetry but preserving lepton number 
can depend linearly on flavor-diagonal parameters
$b_\ka^e$, $b_\ka^\mu$, etc. 
This means that experimental bounds from processes 
preserving lepton number
are typically many orders of magnitude sharper
than bounds involving lepton-number violation.

Consider first spectroscopic studies of muonium $M$, 
which is a $\mu^+$-$e^-$ bound state.
In experiments at RAL and LANL,
precisions of about 20 ppb
have been attained both for the 1S-2S transition
\cite{mu1s2s}
and for the ground-state Zeeman hyperfine transitions
\cite{muonium99}.
However,
we restrict attention here to the latter 
because the hyperfine transition frequencies are much smaller than
the 1S-2S ones,
which implies better absolute energy resolution 
and corresponding sensitivity to CPT and Lorentz violation
\cite{note1}.

The four hyperfine ground states of $M$ can be labeled 
1,2,3,4 in order of decreasing energy. 
The Zeeman hyperfine transitions $\nu_{12}$, $\nu_{34}$
have been measured in a 1.7 T magnetic field
\cite{note2}
with a precision of about 40 Hz ($\sim$ 20 ppb),
and the hyperfine interval has been extracted.

Since electromagnetic transitions in $M$ conserve lepton number,
dominant effects in the standard-model extension
arise from flavor-diagonal terms in Eq.\ \rf{lqed}.
For the case of an antimuon $\mu^+$,
the modified Dirac equation is
\bea
&&\left( i \ga^\la D_\la - m_\mu + a_\la^\mu \ga^\la
- b_\la^\mu \ga_5 \ga^\la
+ \half H_{\ka \la}^\mu \si^{\ka \la} 
\right.
\nonumber \\
&&
\qquad\qquad\qquad
\left.
+ i c_{\ka \la}^\mu \ga^\ka D^\la 
+ i d_{\ka \la}^\mu \ga_5 \ga^\ka D^\la \right) \ps = 0
\quad ,
\label{dirac}
\eea
where $\ps$ is a four-component $\mu^+$ field of 
mass $m_\mu$.
A similar equation exists for the $e^-$,
containing parameters
$a_\la^e$, $b_\la^e$, $H_{\ka \la}^e$, $c_{\ka \la}^e$, 
$d_{\ka \la}^e$.
The associated hamiltonians are found using established procedures
\cite{bkr9798}.
The Coulomb potential in $M$ is $A^\la = (|e|/4 \pi r, \vec 0)$.

The leading-order Lorentz-violating energy shifts in $M$ 
can be obtained from these hamiltonians using
perturbation theory and relativistic two-fermion techniques
\cite{breit}.
For the four Zeeman hyperfine levels in a 1.7 T magnetic field,
we thereby can  determine the corresponding shifts 
$\de\nu_{12}$, $\de\nu_{34}$
in the frequencies $\nu_{12}$, $\nu_{34}$.
We find
\beq
\de \nu_{12} \approx - \de \nu_{34} \approx
-\tilde b_3^\mu/\pi
\quad ,
\label{nu1234}
\eeq
where
$\tilde b_3^\mu \equiv b_3^\mu + d_{30}^\mu m_\mu + H_{12}^\mu$.
Although in a weak or zero field 
\cite{note}
the results would depend on a combination of 
both muon and electron parameters for Lorentz violation,
only the muon parameters appear in Eq.\ \rf{nu1234}
because in a 1.7 T field 
the relevant transitions essentially involve pure muon-spin flips.
Note that subleading-order Lorentz-violating effects 
are further suppressed 
by powers of $\al$ or $\mu_\mu B/m_\mu \simeq 5 \times 10^{-15}$
and can therefore be neglected.

Since the laboratory frame rotates with the Earth, 
and since the frequency shifts \rf{nu1234}
depend on spatial components of the parameters 
for CPT and Lorentz violation, 
the frequencies $\nu_{12}$, $\nu_{34}$
oscillate about a mean value
with frequency equal to the Earth's sidereal frequency
$\Om \simeq 2 \pi$/(23 h 56 m).
Note that no signal of this type emerges at any perturbative order
in the usual standard model without Lorentz violation.
Also, the anticorrelation of the variations of 
$\de \nu_{12}$ and $\de \nu_{34}$ 
could help exclude environmental systematic effects
in analyzing real data.

The result \rf{nu1234} could directly be used to place a bound
on CPT and Lorentz violation in the laboratory frame.
However,
for purposes of comparison among experiments
it is much more useful to work with quantities 
defined with respect to a nonrotating frame.
A suitable choice of basis $\{ \hat X, \hat Y, \hat Z \}$
for a nonrotating frame 
is standard celestial equatorial axes,
with the $\hat Z$ direction oriented along the
the Earth's rotational north pole
\cite{kl99}.
Then,
the laboratory-frame quantity $\tilde b_3^\mu$
can be written as  
\beq
\tilde b_3^\mu = \tilde b_Z^\mu \cos \ch 
+ (\tilde b_X^\mu \cos \Om t + \tilde b_Y^\mu \sin \Om t)\sin \ch 
\quad ,
\label{b3}
\eeq
where the nonrotating-frame quantity 
$\tilde b_J^\mu$ with $J = X,Y,Z$
is defined by 
$\tilde b_J^\mu \equiv b_J^\mu
+ m_\mu d_{J0}^\mu + \fr 1 2 \ep_{JKL} H_{KL}^\mu$,
and where
$\ch$ is the angle between $\hat Z$ 
and the quantization axis
defined by the laboratory magnetic field.

Suppose, 
for definiteness,
that a reanalysis of the data in Ref.\ \cite{muonium99} 
using time stamps on the frequency measurements 
places a bound of 100 Hz on the amplitude of sidereal variations
$\de \nu_{12}$. 
In terms of nonrotating-frame components,
this corresponds to the constraint
\beq
| \sin \ch | ~
\sqrt{(\tilde b_X^\mu)^2 + (\tilde b_Y^\mu)^2}
\lsim 2\times 10^{-22} \, {\rm GeV}
\quad .
\label{bound1}
\eeq
An appropriate dimensionless figure of merit for this result
is the ratio 
$(r^\mu_{\rm hf})_{\rm sidereal}$
of the amplitude of energy variations 
to the relativistic energy of $M$.
The bound \rf{bound1} gives
\beq
(r^\mu_{\rm hf})_{\rm sidereal}
\approx 2\pi |\de \nu_{12}|/m_\mu 
\approx 2 \pi|\de \nu_{34}|/m_\mu
\lsim 4 \times 10^{-21} ,
\label{rhf}
\eeq
which is comparable to the dimensionless ratio 
of the $\mu^+$ mass to the Planck scale $M_P$, 
$m_\mu/M_P \simeq 10^{-21}$.

We consider next measurements of 
the muon anomalous magnetic moment 
\cite{cernmug,farley,muong99}.
The most recent experiment 
\cite{muong99}
measures the angular anomaly frequency $\om_a$,
which is the difference between the spin-precession frequency $\om_s$
and the cyclotron frequency $\om_c$.
This BNL experiment uses relativistic polarized $\mu^+$ 
moving in a constant 1.45 T magnetic field.
The $\mu^+$ have momentum $p \simeq 3.09$ GeV
and `magic' $\ga \simeq 29.3$,
which eliminates the dependence of $\om_a$ 
on the electric field.
Positrons from the decay 
$\mu^+ \rightarrow e^+ + \nu_e + \bar \nu_\mu$
are detected and their decay spectrum is fitted to a
specified time function.  
The anomaly frequency $\om_a$,
which in conventional theory is proportional to $(g-2)/2$,
is measured to about 10 ppm.
An accuracy below 1 ppm is expected in the near future.

In the standard-model extension,
the relativistic hamiltonian for a $\mu^+$ 
with an anomalous magnetic moment 
in a magnetic field $\vec B$ is 
\bea
\hat H 
&=& \ga^0 \vec \ga \cdot \vec \pi
+ m_\mu \left(1 - c_{00}^\mu \right) \ga^0 
+ \half (g-2)\mu_\mu \ga^0 \vec \Si \cdot \vec B
\nonumber \\
&&  
- a_0^\mu -\left(c_{0j}^\mu +  c_{jo}^\mu \right) \pi^j
- \left[ b_0^\mu + (d_{0j}^\mu+d_{j0}^\mu) \pi^j \right] \ga_5
\nonumber \\
&& 
- \left[a_j^\mu+ (c_{jk}^\mu+c_{00}^\mu \de_{jk}) \pi^k \right]
  \ga^0 \ga^j - i H_{0j}^\mu \ga^j
\nonumber \\
&& 
- \left[b_j^\mu+ (d_{jk}^\mu+d_{00}^\mu \de_{jk}) \pi^k \right]
  \ga _5 \ga^0 \ga^j
\nonumber \\
&&  
+ \left[ \half\ve_{jkl}H_{kl}^\mu + md_{j0}^\mu \right]
   \ga _5 \ga^j
   \quad ,
\label{ham}
\eea 
where $\Si^j = \ga_5 \ga^0 \ga^j$,
$\mu_\mu$ is the muon magneton,
and $\vec \pi = \vec p - q \vec A$,
with $q = + |e|$ for $\mu^+$.
This hamiltonian contains no terms
that provide leading-order corrections
to the $g$ factors for $\mu^+$ or $\mu^-$.
Instead,
the dominant sensitivity to CPT violation 
results from the sensitivity to small frequency shifts
associated with the spin precession.
The conventional figure of merit $r^\mu_g$ therefore is zero
at leading order despite the presence of explicit CPT violation,
which means alternative figures of merit are needed
\cite{bkr9798}.

A Foldy-Wouthuysen transformation 
\cite{fw}
can be used to convert the hamiltonian $\hat H$
to another hamiltonian $\hat H^\prime$ 
in which the $2\times 2$ off-diagonal blocks 
contain only first-order terms
in the magnetic field $\vec B$ 
\cite{mc55}
and in the parameters for CPT and Lorentz violation.
We find 
$\hat H^\prime = 
\exp (\ga^0 \ga_5 \ph) \hat H \exp (-\ga^0 \ga_5 \ph)$,
with 
$\tan 2 \ph = |\vec \Si \cdot \vec \pi |/m_\mu$
and $|\vec \Si \cdot \vec \pi |^2
= \vec \pi^2 - q \vec \Si \cdot \vec B$.
The off-diagonal blocks in $\hat H^\prime$ 
are irrelevant at leading order 
since here they produce effects that are at least quadratic
in small parameters.

The upper-left $2 \times 2$ block of $\hat H^\prime$
is the relevant relativistic hamiltonian for the $\mu^+$ 
in the laboratory frame
\cite{note3}.
It has the form
\beq
\hat H^\prime = \cE_0 + \cE_1 + \half \vec \si \cdot 
( \vec \om_{s,0} + f_1 \vec \be + \vec f_2 )
\quad ,
\label{FWham}
\eeq
where $\cE_0 = \ga m$
and $\ga = (1-\be^2)^{-1/2}$
with 3-velocity $\vec \be$.
The term $\cE_1$ contains irrelevant spin-independent corrections.
The quantity 
$\vec \om_{s,0} = (g-2 + 2/ \ga) \mu_\mu \vec B$
is the usual spin-precession frequency.
The term $f_1 \vec \be$ is proportional to $\vec \be$,
and its contributions average to zero
since the detectors in the $(g-2)$ experiments
are spread around the ring and their data are summed.
The term $\vec f_2$ depends on 
the parameters for CPT and Lorentz violation 
and partially on $\vec \be$,
but again only the $\hat\be$-independent terms are relevant here.

The spin-precession frequency $\om_s$ is calculated as 
$ \fr {d \vec \si} {dt} =
i [\hat H^\prime, \vec \si] = \vec \om_s \times \vec \si$.
Since the detectors are in the $\hat x$-$\hat y$ plane
in the laboratory frame,
only the vertical component $\om_s$ is measured.
Substituting for $\hat H^\prime$ 
and keeping only the velocity-independent terms along
the $\hat z$ direction gives for $\mu^+$ the result
$\om_s \approx \om_{s,0} + 2 \mathaccent 20 b_3^\mu$,
where $\mathaccent 20 b_3^\mu \equiv 
b_3^\mu/\ga + m_\mu d_{30}^\mu + H_{12}^\mu$.
Note that $\mathaccent 20 b_3^\mu$
reduces to $\tilde b_3^\mu$
in the nonrelativistic limit
\cite{note4}.

The cyclotron frequency $\om_c$ is obtained from 
$[\hat H^\prime, \vec r] = \vec \pi/\cE_0$,
which contains a term $\vec \om_c \times \vec r$.
However,
no leading-order corrections 
to the usual cyclotron frequency appear:
$\om_c \approx \om_{c,0} = 2 \mu_\mu B/\ga$.
Subleading-order terms do in fact contribute
but are of lower order than those in $\om_s$
and therefore can be ignored.

Combining the above results
and converting to the nonrotating frame as in Eq.\ \rf{b3},
we find the correction to the $\mu^+$ anomaly frequency 
$\om_a = \om_s - \om_c$
due to CPT and Lorentz violation is 
\beq
\de \om_a^{\mu^+} \approx 
2 \mathaccent 20 b_Z^\mu \cos \ch 
+ 2 (\mathaccent 20 b_X^\mu \cos \Om t
+ \mathaccent 20 b_Y^\mu \sin \Om t) \sin \ch ,
\label{delwa}
\eeq
where $\ch$ is now the colatitude of the experiment.
The corresponding expression  
$\de \om_a^{\mu^-}$ for $\mu^-$ 
is obtained by the substitution
$b_J^\mu \rightarrow - b_J^\mu$
in the expressions for
$\mathaccent 20 b_X, \mathaccent 20 b_Y, \mathaccent 20 b_Z$.

These results suggest two interesting types of experimental signal.
The first involves the difference
$\De \om_a^\mu \equiv \de \om_a^{\mu^+} - \de \om_a^{\mu^-}$,
which is
$\De \om_a^\mu \approx 4 b_3^\mu/\ga$
in the laboratory frame
\cite{note5}.
It is impractical to measure $g-2$ for both $\mu^+$
and $\mu^-$ simultaneously,
so instead one can directly consider the time-averaged  
difference $\overline{\De \om_a^\mu}$.
In the nonrotating frame,
\beq
\overline{\De \om_a^\mu} \approx \fr 4 {\ga} b_Z^\mu \cos \ch
\quad .
\label{delwaav}
\eeq
An appropriate figure of merit $r^\mu_{\De\om_a}$ here 
is the relative energy difference between $\mu^+$
and $\mu^-$ caused by their different spin precessions: 
\beq
r^\mu_{\De\om_a} \approx 
\overline{\De \om_a^\mu} /m_\mu 
\quad.
\label{rwa}
\eeq

The CERN $g-2$ experiments compared 
average $\mu^+$ and $\mu^-$ anomaly frequencies,
finding 
\cite{cernmug}
$\overline{\De \om_a^\mu} /2\pi \simeq 5 \pm 3$ Hz.
This gives a value of $r^\mu_{\De\om_a}$ on the order of
$2 \times 10^{-22}$,
corresponding to $b_Z^\mu \simeq (2\pm 1) \times 10^{-22}$ GeV.
A subsequent measurement at BNL 
\cite{muong99}
provides a $\mu^+$ result within one standard deviation 
of the CERN $\mu^-$ result.
If the BNL experiment eventually limits the
frequency difference to 1 ppm,
it would provide a sensitivity
at the level of $r^\mu_{\De\om_a} \lsim 10^{-23}$,
corresponding to $b_Z^\mu \lsim 10^{-23}$ GeV.

The second interesting type of experimental signal
involves sidereal variations in the anomaly frequency.
It can be studied using $\mu^+$ alone,
in which case time stamps on frequency measurements 
would permit a bound on sidereal variations of $\om_a^{\mu^+}$.
An appropriate figure of merit 
$(r_{\om_a}^\mu)_{\rm sidereal}$ 
is the relative size
of the amplitude of energy variations 
compared to the total energy.
Assuming a precision of 1 ppm,
we estimate an attainable bound of 
\beq
(r_{\om_a}^\mu)_{\rm sidereal} \approx
|\de \om_a^{\mu^+}|/m_\mu
\lsim 10^{-23} 
\quad .
\label{romadiurnal}
\eeq
The associated bound on parameters in
the nonrotating frame is
\beq
| \sin \ch | ~
\sqrt{(\mathaccent 20 b_X^\mu)^2 
+ (\mathaccent 20 b_Y^\mu)^2}
\lsim 5 \times 10^{-25} \, {\rm GeV}
\quad ,
\label{bound2}
\eeq
which again represents sensitivity to the Planck scale.
Note that this test 
involves different sensitivity to CPT violation 
than the previous one:
the two figures of merit 
$r^\mu_{\De\om_a}$,
$(r_{\om_a}^\mu)_{\rm sidereal}$ 
depend on independent components of parameters for  
CPT and Lorentz violation.

In addition to effects in flavor-diagonal processes,
off-diagonal terms of the type in Eq.\ \rf{amatrix} 
arising in the standard-model extension allow
Lorentz-violating contributions to flavor-changing processes.
For example,
precision searches have been performed 
for the radiative muon decay $\mu \rightarrow e \ga$,
which has a branching ratio below $5 \times 10^{-11}$
\cite{murad}.
This decay has previously been analyzed 
using a CPT- and rotation-invariant model 
with Lorentz and lepton-number violation
that involves terms equivalent (up to field renormalizations) 
to those of the form $c_{00}^{e\mu}$ and $d_{00}^{e\mu}$
in Eq.\ \rf{lqed}
\cite{cg99}.
The results of this analysis indicate
that combinations of the dimensionless parameters
$c_{00}^{e\mu}$ and $d_{00}^{e\mu}$
are bounded at the level of about $10^{-12}$
by rest-frame muon decays or 
by muon lifetime measurements in the CERN $g-2$ experiments,
and at the level of about $10^{-19}$
by constraints from horizontal air showers of cosmic-ray muons.
As expected from the discussion following Eq.\ \rf{amatrix},
these bounds are several orders of magnitude weaker than those 
from lepton-number preserving processes.
An extension of this analysis to include all types of term 
in Eq.\ \rf{lqed} would provide the best existing bounds
on the flavor-nondiagonal parameters in the electron-muon sector
of the standard-model extension.
Useful constraints on these parameters 
could also be extracted from other future experiments.
These include the proposed tests for muon-electron conversion 
\cite{muco},
which have an estimated sensitivity 
to the process $\mu^- + N \rightarrow e^- + N$
of $2 \times 10^{-17}$,
and the various precision tests that might be envisaged
at a future muon collider.

\smallskip

We thank D.\ Hertzog, K.\ Jungmann, and B.L.\ Roberts for discussion.
This work is supported in part by the U.S.\ D.O.E.\
under grant number DE-FG02-91ER40661 and by the N.S.F.\ 
under grant number PHY-9801869.

\end{multicols}

\begin{references}
\bibitem{note1}
In hydrogen and antihydrogen,
a 1S-2S frequency resolution of about 1 mHz might be attainable
in principle.
However,
leading-order sensitivity to CPT and Lorentz violation
requires magnetic mixing of the 1S and 2S spin states,
which introduces experimental difficulties associated with
Zeeman field broadening
\cite{bkr99}.
Similar issues would be relevant for experiments
measuring the 1S-2S transition in muonium.
\bibitem{note2}
In these and $g-2$ experiments,
the magnetic field is determined 
in terms of the proton NMR frequency.
In principle,
this could introduce additional Lorentz- and CPT-violating
effects involving the proton.
However,
clock-comparison experiments
\cite{cctests,kl99}
suggest any such effects are too small 
to affect the muon tests considered here.
\bibitem{note}
Even in a zero field,
the hyperfine interval is split.
\bibitem{note3}
An expression for the full hamiltonian $\hat H^\prime$ 
in the nonrelativistic limit is presented in 
Ref.\ \cite{kl99}.
\bibitem{note4}
Intuition about this result can be obtained by considering
an instantaneous observer Lorentz boost 
from the muon rest frame to the laboratory.
The frequency $\om_s^{\mu}$ scales as $\ga$,
$b_3^\mu$ is unaffected since it is perpendicular to the boost,
and $d_{30}^\mu$ and $H_{12}^\mu$
each scale as $\ga$.
The net effect produces the 
stated $\ga$-dependence in the definition of 
$\mathaccent 20 b_3^\mu$.
\bibitem{note5}
Note that $\De \om_a^\mu$ involves only the CPT-violating
parameter $b_3^\mu$ instead of the combination 
$\mathaccent 20 b_3^\mu$.
The CPT-preserving parameters $d_{30}^\mu$ and
$H_{12}^\mu$ cancel in particle-antiparticle comparisons.
\end{references}
\end{document}